\documentclass[aps,prl,superscriptaddress,twocolumn,showpacs,amsmath,amssymb,reprint,floatfix]{revtex4-1}

\usepackage{graphicx}
\usepackage{dcolumn}
\usepackage{bm}
\usepackage{lipsum}
\usepackage{chemformula}
\usepackage[T1]{fontenc}
\usepackage[colorlinks=true, allcolors=blue]{hyperref}
\usepackage{amsfonts}
\usepackage{algorithm2e}
\usepackage[utf8]{inputenc}
\usepackage[T1]{fontenc}
\usepackage{mathptmx}

\begin{document}

\title{Room temperature two-dimensional antiferromagnetic Weyl semimetal CrO with giant spin-splitting and spin-momentum locked transport}

\author{Xin Chen}
\affiliation{Department of Physics and Astronomy, Uppsala University, Box 516,
751\,20 Uppsala, Sweden}
\author{Duo Wang}
\affiliation{Department of Physics and Astronomy, Uppsala University, Box 516,
751\,20 Uppsala, Sweden}
\author{Linyang Li}
\email{linyang.li@hebut.edu.cn}
\affiliation{School of Science, Hebei University of Technology, Tianjin 300401, China}
\author{Biplab Sanyal}
\email{biplab.sanyal@physics.uu.se}
\affiliation{Department of Physics and Astronomy, Uppsala University, Box 516,
751\,20 Uppsala, Sweden}
\date{\today}

\begin{abstract}
Giant spin-splitting was recently predicted in collinear antiferromagnetic materials with a specific class of magnetic space group. In this work, we have predicted a two-dimensional (2D) antiferromagnetic Weyl semimetal (WS), CrO with large spin-split band structure, spin-momentum locked transport properties and high Néel temperature. It has two pairs of spin-polarized Weyl points at the Fermi level. By manipulating the position of the Weyl points with strain, four different antiferromagnetic spintronic states can be achieved: WSs with two spin-polarized transport channels (STCs), WSs with single STC, semiconductors with two STCs, and semiconductors with single STC. Based on these properties, a new avenue in spintronics with 2D collinear antiferromagnets is proposed.
\end{abstract}

\maketitle

Spin-polarized currents with spin-dependent conductivities are one of the prerequisites of spintronic devices. \cite{Wolf1488} In this regard, the giant magnetoresistance (GMR) effect \cite{PhysRevLett.61.2472,PhysRevB.39.4828} and spin-transfer torque (STT) \cite{PhysRevB.39.6995,RALPH20081190} have been firmly established and are employed in many fields, such as magnetic random access memories and magnetic sensors \cite{Khvalkovskiy_2013}. Another primary method is using the spin-orbital coupling (SOC) induced spin Hall effects (SHEs) and Rashba-Edelstein effects, which give rise to damping-like and field-like spin-orbital torques (SOTs) \cite{Miron2011,Liu555}. The SOTs can be used for controlling the magnetization in next-generation low-power magnetic random access memories \cite{7008441}. However, a sufficient large SOC can only be achieved in materials that contain heavy atoms, which are often fragile due to the weak chemical bonds \cite{doi:10.1002/pssb.19680250130}. 

Though spintronics was first discovered in ferromagnetic (FM) materials, spin-polarized currents in antiferromagnetic (AFM) materials have also attracted enormous attention \cite{Jungwirth2016,Wadley587,Lavrijsen2019,10.1093/nsr/nww026,RevModPhys.90.015005}. Superior to FM spintronic devices, AFM materials do not have any net magnetic moment and are robust to external magnetic perturbation \cite{doi:10.1002/adma.201102555}. In addition, AFM materials have been found to have ultra-high dynamic speed, which allows high-speed device operation \cite{Jungwirth2018}. However, in collinear AFM systems, due to the absence of spin-splitting (SS) in the band structures, it is often hard to achieve spin-polarized currents \cite{PhysRevLett.119.187204}.

While a number of half-metallic FMs with 100 \% spin-polarized current has been realized, half-metallic antiferromagnets (HFAFMs) are not so commonly found. The concept of HFAFM was first proposed in 1995 by Leuken, and Groot \cite{PhysRevLett.74.1171}. They predicted that carefully designed Heusler alloys could be half-metallic without net magnetization. There is also some experimental evidence of this exciting phenomenon in recent years \cite{PhysRevLett.112.027201,Nayak2015}, but all of them are in alloys.

Recently, the spintronic phenomenon of momentum-dependent spin-splitting (SS) in some collinear AFM crystals and the induced spontaneous Hall effect have attracted much interest \cite{smejkal2020crystal}. The SS is only originated from the simple AFM order, and heavy atoms are thus not needed. Several materials have been predicted to have this SS, such as RuO$_{2}$ \cite{smejkal2020crystal,feng2020observation}, some organic AFMs \cite{Naka2019}, MnF$_{2}$ \cite{PhysRevB.102.014422}, and some GdFeO3-type perovskites \cite{naka2020perovskite}. Moreover, Yuan et al. have also paved a way to identify and design the AFM materials with this feature by using their magnetic space group \cite{PhysRevB.102.014422,yuan2020prediction}.

Weyl semimetals, as a new condensed matter phase, have attracted a lot of research interest. Different from the Dirac points in graphene and many spin-gapless semiconductors (SGSs) \cite{PhysRevLett.100.156404,PhysRevLett.109.237207}, Weyl points are two-fold degenerate and do not open up a gap under SOC \cite{PhysRevLett.95.226801,PhysRevB.100.064408}. Many materials have been reported to be 3D Weyl semimetals, while 2D Weyl semimetals are rarely reported. You et al. have recently reported that monolayer PtCl$_{3}$ is a 2D half-metallic Weyl semimetal: without considering the spin-orbital coupling, there is a Dirac cone at the Fermi level, while under SOC, the Dirac cone is split into two Weyl points \cite{PhysRevB.100.064408}. Another example is non-magnetic spin–valley-coupled Dirac semimetals (svc-DSMs) reported by Liu et al., in which the Weyl points are originated from SOC-induced giant SS under inversion symmetry breaking \cite{C8MH01588K}. So far, 2D Weyl semimetals originated in the absence of SOC are not reported.

In this work, we have predicted a collinear AFM material with giant momentum-dependent SS, monolayer CrO with square symmetry (namely s-CrO). Interestingly, s-CrO is also a semimetal with two pairs of spin-polarized Weyl points at the Fermi level. Due to the anisotropy of the Weyl points, spin-polarized transport can be achieved in s-CrO. A critical feature of Weyl points is that when a pair of Weyl points merge, both of them are annihilated. By applying uniaxial strain, one can move and annihilate a pair of fully spin-polarized Weyl points and open up a gap at one spin-polarized transport channel (STC) while preserving the Weyl point in the other STC, resulting in a half-metallic AFM semimetal. With biaxial strain, an AFM semiconductor with only one STC (half-semiconductor) can also be achieved.

We performed first-principles density functional calculations by using Vienna ab initio Simulation Package (VASP)\cite{002230939500355X,PhysRevB.54.11169} code. The exchange-correlation-functional was approximated by local spin density approximation. To account for electron correlation, we have considered Hubbard U formalism for correlated Cr-d orbitals. The energy cutoff of the plane wave basis was chosen as 520 eV, and the Brillouin zone (BZ) was sampled by Monkhorst-Pack grid denser than  $2 \pi \times 0.02$ \AA$^{-1}$.  The out-of-plane interaction was avoided by taking a vacuum of more than 20 \AA. The tolerance for energy convergence was set to be less than 10$^{-7}$ eV. We optimized the structures until the force on each atom became smaller than 0.001 eV/\AA. For examining the dynamical stability of structures, phonon spectra were computed using the PHONOPY code \cite{phonopy}. The interatomic exchange parameters were computed using the full-potential linear muffin-tin orbital (FP-LMTO) code RSPt \cite{wills2010full}. This is achieved by mapping the magnetic excitations onto the Heisenberg Hamiltonian:
\begin{equation}
    \hat{H}=-\sum_{i \neq j}J_{ij} \Vec{e_{i}} \cdot \Vec{e_{j}},
\end{equation}
where $J_{ij}$ is the interatomic exchange interaction between the two magnetic moments at sites $i$ and $j$, $\Vec{e_{i}}$ is a unit vector along the magnetization direction at the site $i$. Finally, we used the extracted $J_{ij}$ to calculate the magnetic ordering temperature by means of a classical Monte Carlo (MC) algorithm for the solution of the Heisenberg Hamiltonian as implemented in the Uppsala Atomistic Spin Dynamics (UppASD) code \cite{Skubic_2008}. For this purpose, a $100 \times 100 \times 1$ (20000 atoms) cell was considered. The adiabatic magnon spectra were calculated for a considered magnetic ground state from the $J_{ij}$'s and magnetic anisotropy energy (MAE).

To determine the proper U value, we first computed the magnetic moment of s-CrO employing Heyd-Scuseria-Ernzerhof (HSE) hybrid functional within the framework of HSE06 \cite{Krukau2006}, and then benchmarked with the computed magnetic moments of s-CrO by using different U values in LSDA+U calculations. As shown in Fig. S1 in the Supplemental Material (SM) \cite{SuppMater}, the LSDA+U result reproduces the HSE result when U=3.55 eV. For further calculations, this U value was used.

\begin{figure}[htbp]
\includegraphics[scale=1.0]{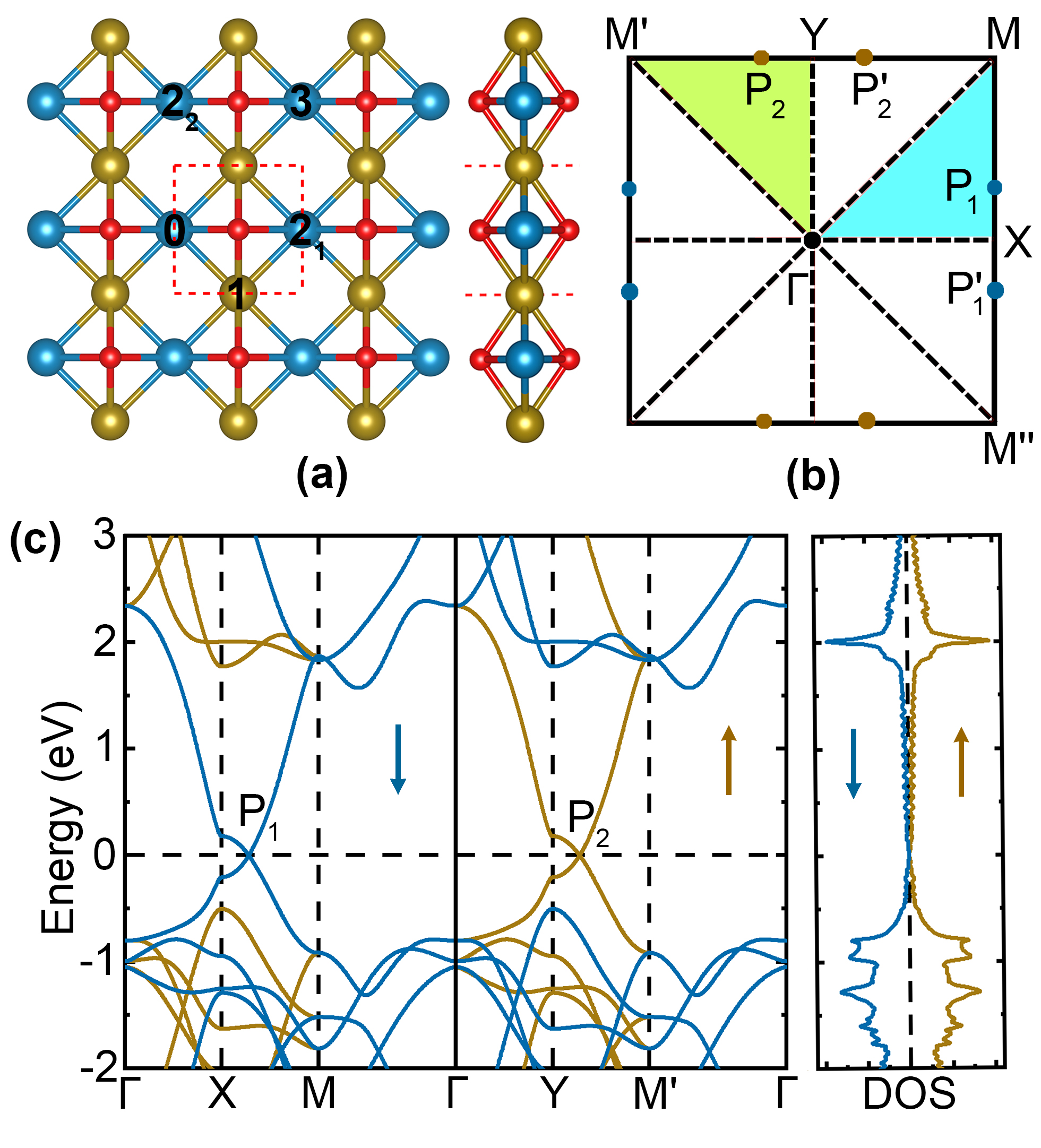} 
\caption{(a) The top and side views of the structure of AFM s-CrO. The red balls are O atoms, and the green and blue balls are Cr atoms with spin-up and spin-down magnetic polarization, respectively. The first, second, and third nearest magnetic atoms of the atom 0 are indicated by the respective numbers. (b) The first BZ of AFM s-CrO, with high symmetry points and all Weyl points are shown. The green and blue areas are the two irreducible parts of the first BZ. (c) The band structure and total DOS of AFM s-CrO are shown. The spin-up and spin-down channels are depicted in green and blue. The Fermi level is set as 0 eV.
}
\label{figure1}
\end{figure}

Our calculated structure and magnetic configuration are shown in Fig. \ref{figure1} (a). The unit cell contains two O atoms and two antiferromagnetically coupled Cr atoms, and the lattice parameter is about 3.37 \AA. All Cr atoms are in one plane, sandwiched by two O atomic layers. The AFM order (labeled as AFM1) of s-CrO is determined by comparing the energies of the structure with three different magnetic configurations, as shown in Fig. S2 (a) in the SM \cite{SuppMater}. The value of the energy difference between FM and AFM configurations is affected by U values, but the magnetic order AFM1 is always the lowest energy magnetic configuration, as shown in Fig. S2 (b) in the SM \cite{SuppMater}. With or without considering the AFM order, the primary cell is equivalent. The magnetic space group of s-CrO is P4$^{\prime}$/mmm$^{\prime}$, containing 16 symmetry operators, half of which are unitary. With the above two features, s-CrO fully satisfies the requirement of having giant momentum-dependent SS in the absence of SOC, according to the recent work of Yuan et al. \cite{PhysRevB.102.014422}. The first Brillouin zone (BZ) is shown in Fig. \ref{figure1} (b). One of the irreducible parts of the BZ is the triangular $\Gamma$-X-M. By a rotation of 90 degrees and a time-reversal operation, we can get the other irreducible part of the first BZ, the triangular $\Gamma$-X-M.

\begin{figure}[htbp]
\includegraphics[scale=1.0]{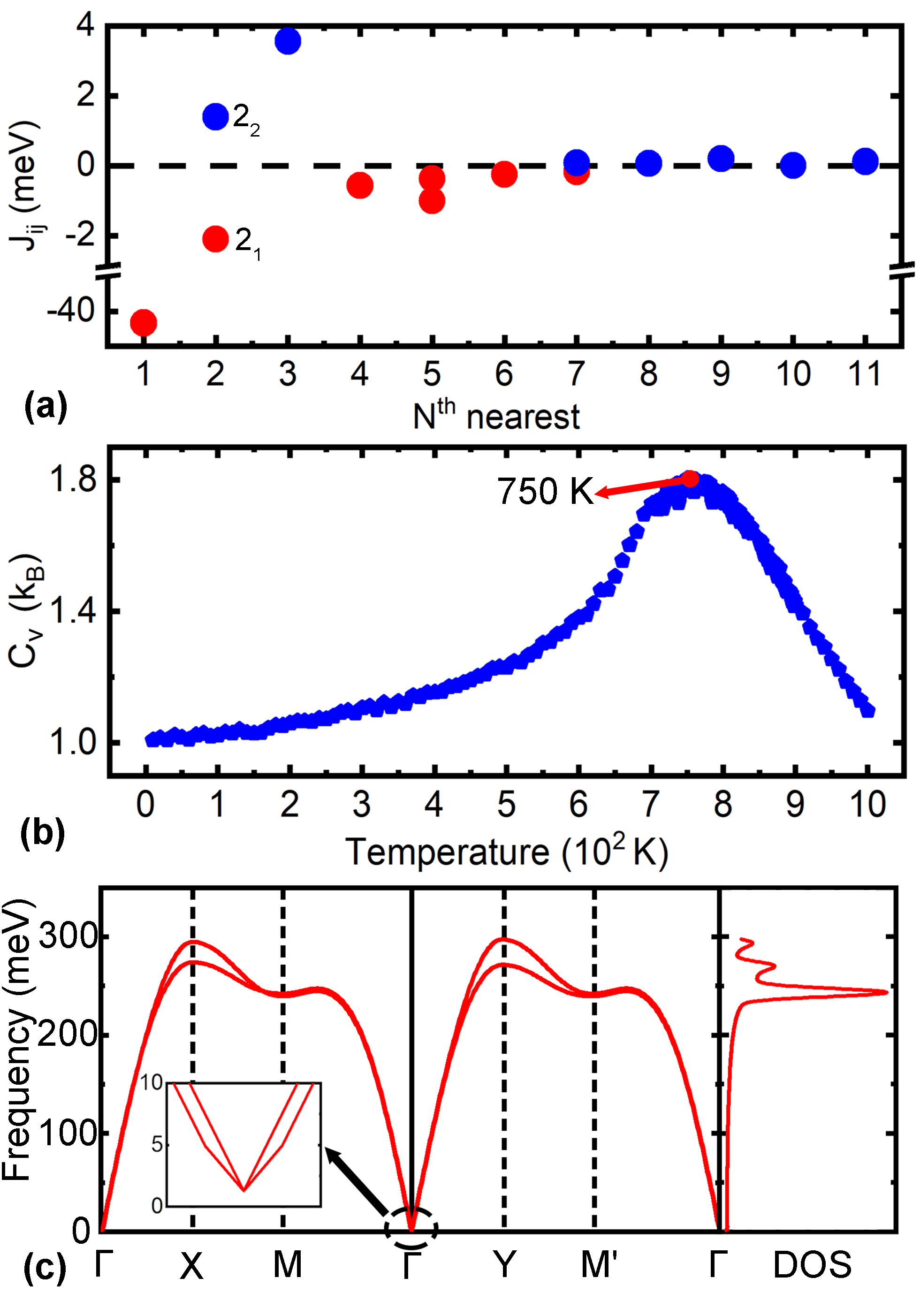} 
\caption{(a) The Exchange parameters between a Cr atom and its N$^{th}$ nearest neighbor Cr atoms. The positive (ferromagnetic) and negative (antiferromagnetic) values are illustrated by blue and red points, respectively. (b) The Monte Carlo results of the specific heat as a function of temperature. (c) The magnon dispersion spectra and corresponding DOS of monolayer CrO.}
\label{figure2}
\end{figure}

We have confirmed the dynamical stability of AFM s-CrO by computing its phonon spectrum and performing ab initio Born-Oppenheimer molecular dynamics (BOMD) simulations. It is observed that in the phonon spectrum of AFM s-CrO, there are no imaginary frequencies. In the BOMD simulation at 300 K, the energies are kept stable, and the geometric features are preserved. The computational details are in Part III of the SM \cite{SuppMater}.

The magnetic easy-axis is investigated by comparing the energies of the system with different spin axis setting by considering spin-orbit coupling in the Hamiltonian. It is found that the easy-axis of AFM s-CrO is in-plane, and the magnetic anisotropy is very weak in the easy-plane. The MAE computed by $E_{MAE}= E_{100} - E_{001}$ is 0.13 meV$/$Cr atom. As shown in Fig. \ref{figure2} (a), in AFM s-CrO, the exchange parameters decrease fast as the distance between two Cr moments increases. The first three nearest neighbor magnetic couplings play the essential role setting the magnetic long-range order and the more distant ones are almost zero. The nearest neighbor exchange coupling has the strongest AFM contribution (40.3 meV). The exchange energy is one order bigger than the others. Interestingly, if compared to the first nearest coupling of bcc Cr (20.4 meV AFM coupling) \cite{10.1103/physrevlett.116.217202}, the first nearest neighbor coupling in CrO is two times more significant. It is related to the short Cr-Cr bond: in monolayer CrO, the Cr-Cr bond length is 2.38 \AA, about 5 $\%$ shorter than that in bcc Cr (2.50 \AA). The strong AFM coupling in monolayer CrO indicates high magnetic ordering temperature, which will be shown later. There are two different chemical environments for the second nearest neighbors, the one with two Cr atoms connected by an O atom ($2_1$ in Fig. \ref{figure1} (a)) coupled antiferromagnetically, and another one with Cr atoms connected by another Cr atom ($2_2$ in Fig. \ref{figure1} (a)) coupled ferromagnetically. The first coupling can be explained by Goodenough-Kanamori-Anderson rules \cite{10.1103/physrev.100.564, 10.1016/0022-3697(59)90061-7, 10.1143/PTP.17.197} quite well: a nearly 180$^{\circ}$ superexchange of two magnetic ions with partially filled d shells is AFM if the virtual electron transfer is between the overlapping orbitals that are each half-filled. In comparison, FM couplings occur at second and third nearest neighbors is mainly because of the robust first nearest neighbor AFM coupling.

By using our calculated interatomic exchange parameters and MAE, we calculated the magnetic ordering temperature and adiabatic magnon spectra. From the calculated specific heat as a function of temperature, as shown in Fig. \ref{figure2} (b), the magnetic ordering temperature is estimated to be 750 K, which indicates that this magnetic configuration is stable at room temperature. In the magnon spectra (as shown in Fig. 
\ref{figure2} (c)), there are two branches, which are correspond to two magnetic ions in the unit cell. As we can see from the inset, there is a energy gap at the $\Gamma$ point, which corresponds to the magnetic anisotropy along 100 direction. The gap is 1.28 meV, which also indicates the weak MAE in the system. Besides that, this MAE also induces an interesting difference in the magnon spectra, which are shown in two Brillioun parts: there is a cross point occurring in the X-M direction, but not in the Y-M$^{\prime}$ direction. In addition, we compared this magnon spectra to another one, which does not include SOC, as shown in Fig. S4 in the SM \cite{SuppMater}. In the latter case, two branches along both M-$\Gamma$ and M$^{\prime}$-$\Gamma$ are degenerate. The spin stiffness constant is in good agreement with the magnon spectra for a AFM material.

The electronic band structure and total density of states (DOS) of AFM s-CrO are shown in Fig. \ref{figure1} (c). In the spin-up channel, the valence band maximum (VBM) and the conduction band minimum (CBM) meet at the Fermi level on the line of X-M. In the spin-down channel, the VBM and CBM also meet at the Fermi level, but on the line of Y-M$^\prime$. According to the magnetic space group, we can find other band-crossings in the first BZ, as shown in Fig. \ref{figure1} (b). The dispersion relationship of the bands in the first BZ near the Fermi level can be intuitively seen in Fig. S5 in the SM \cite{SuppMater}. The electronic band spectra in both spin channels show that AFM s-CrO is spin-gapless. Moreover, protected by the symmetry, the total DOS in the spin-up channel is equal to that in the spin-down channel, which results in zero net magnetization. The origin of the band-crossing is investigated by computing the atomic orbital projected density of states (PDOS), as shown in Fig. S6 in the SM \cite{SuppMater}. It is found that the bands near the Fermi level are dominated by the $d_{x^{2}-y^{2}}$ atomic orbital of Cr atoms and the $p_{z}$ atomic orbital of O atoms. 

With the SOC effect taken into account, there is no band gap opening at the Weyl point, as shown in Fig. S7 in the SM \cite{SuppMater}, which shows that the band-crossing points are indeed two pairs of Weyl points \cite{PhysRevB.100.064408}. One pair of the Weyl points is in the spin-down channel, i.e., P$_{1}$ and P$_{1}^{\prime}$ in Fig. \ref{figure1}. The other pair of Weyl points, i.e., P$_{2}$ and P$_{2}^{\prime}$, is in the spin-up channel. 

\begin{figure}[htbp]
\includegraphics[scale=1]{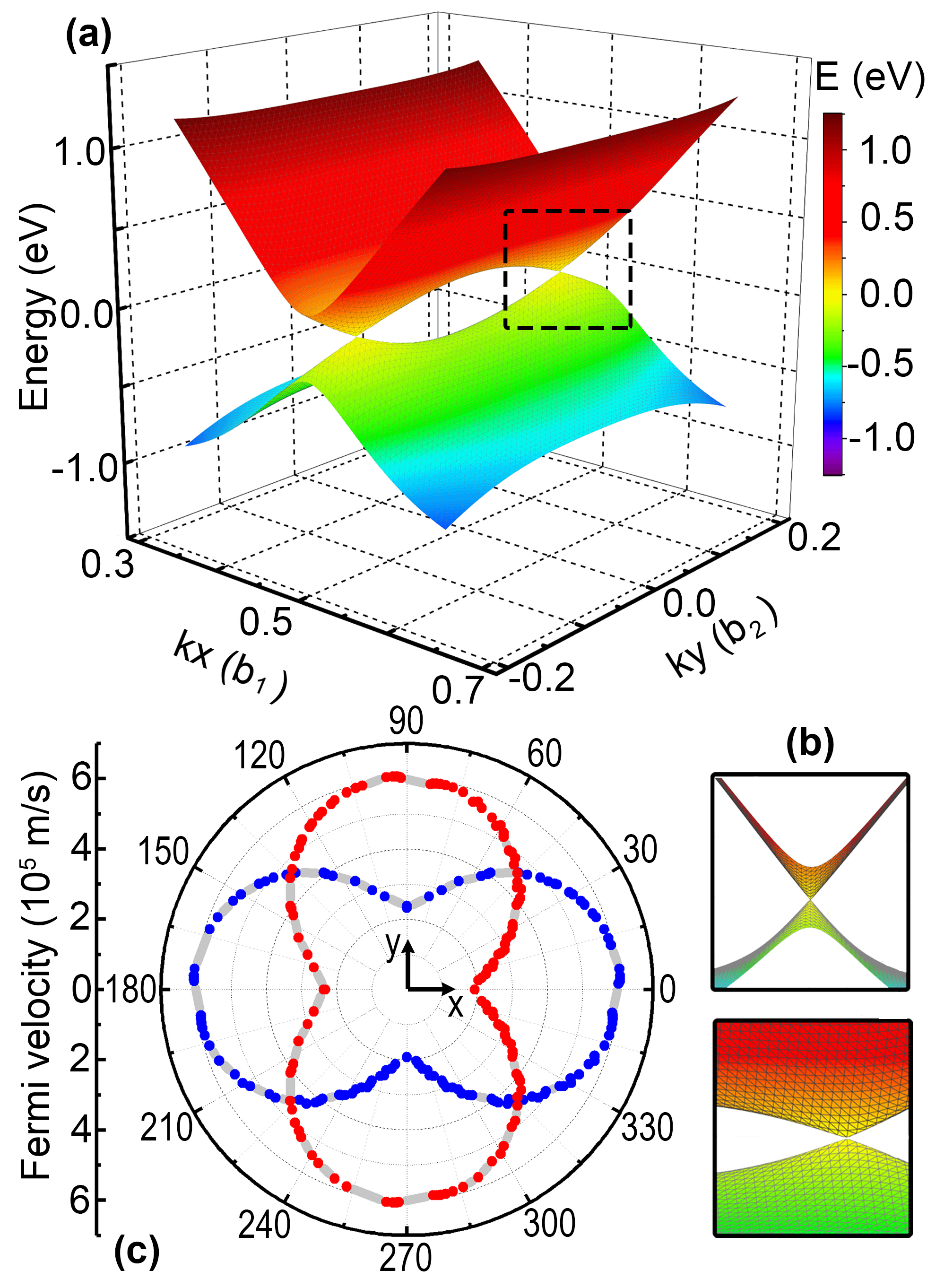}
\caption{(a) The 3D bands of AFM s-CrO near the Weyl points P$_{1}$ (in the dashed rectangle) and P$_{1}^{\prime}$. The enlarged side views along the y-direction and the x-direction are shown in (b). (c) The direction-dependent Fermi-velocities of Weyl points P$_{1}$ (spin-down channel, blue points) and P$_{2}$ (spin-up channel, red points). Note that the red points are obtained by using a 90-degree rotation operator.}
\label{figure3}
\end{figure}
\begin{figure*}[hbtp]
\includegraphics[scale=1]{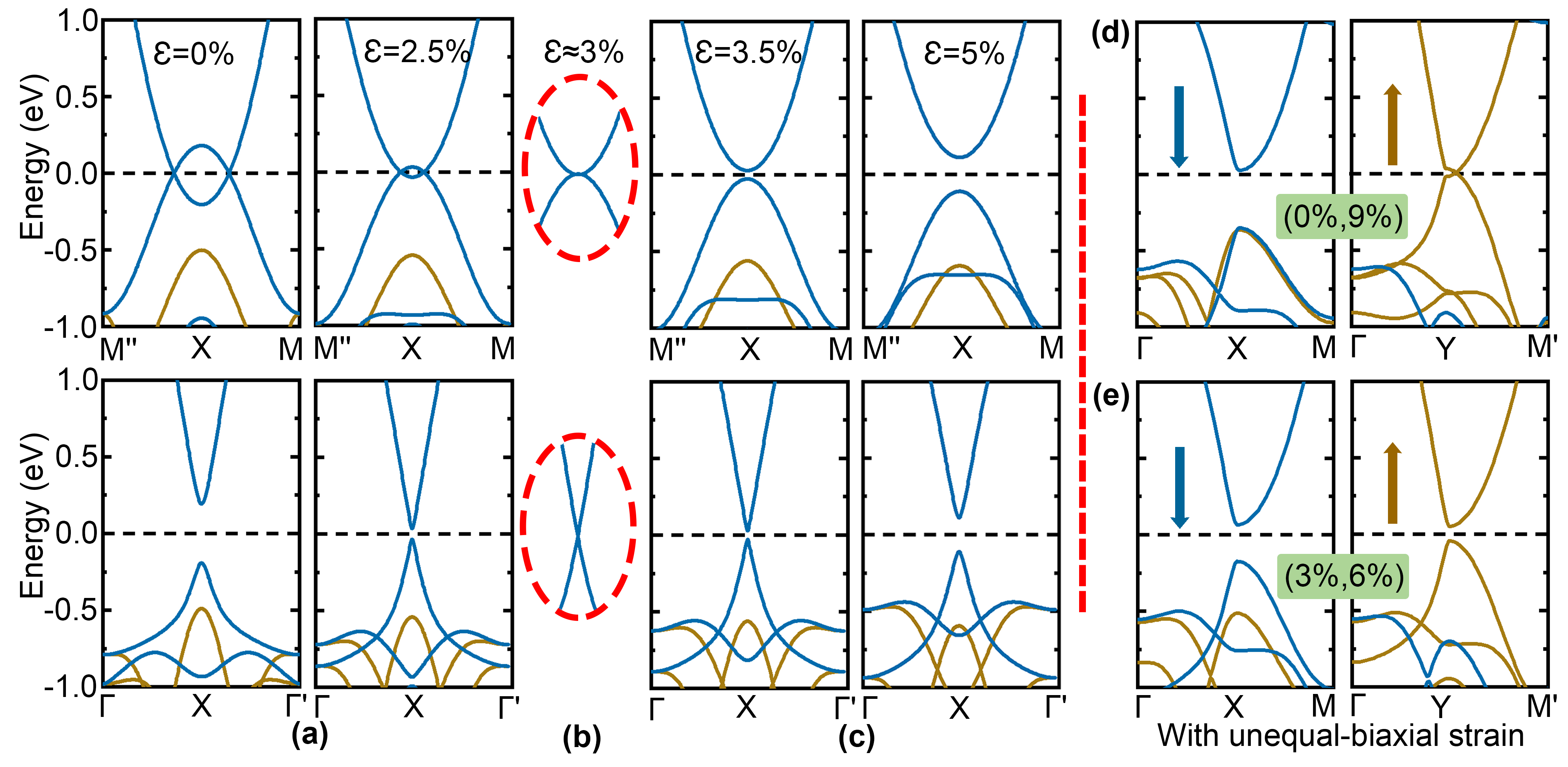}
\caption{Strain induced evolution from (a) a Weyl semimetal passing through (b) a spin-gapless semiconducting state to (c) a direct gap semiconductor. The up-panel shows the bands along M$^{\prime}-X-M$ (in the y-direction). The down-panel shows the bands along $\Gamma-X-\Gamma^{\prime}$ (in the x-direction), in which $\Gamma^{\prime}=(\mathbf{b}_{1},0,0)$. (d) and (e) are the electronic bands under unequal-biaxial tensile strains $(\epsilon_{x}, \epsilon_{y})$ of $(0\%, 9\%)$ and $(3\%, 6\%)$, respectively.}
\label{figure4}
\end{figure*}

We computed the 3D bands near P$_{1}$ and P$_{1}^{\prime}$, as shown in Fig. \ref{figure3} (a). It can be observed that the Weyl points are highly anisotropic. As shown in Fig. \ref{figure3} (b), in the x-direction, the Weyl point has symmetric linear dispersion. In contrast, in the y-direction, the linear dispersion is not symmetric. Moreover, the slope of the bands in the x-direction is much larger than that in the y-direction, giving rise to highly anisotropic transport properties in the spin-down channel.

The anisotropic transport properties are investigated by computing the direction-dependent Fermi velocities at the Weyl points in two spin-channels. As shown in Fig. \ref{figure3} (c), in the spin-down channel, the Fermi-velocity of Weyl point P$_{1}$ reaches its maximum in the x-direction and reaches its minimum in the y-direction. The maximum of the Fermi velocity is about $6 \times 10^{5}$ m/s, comparable to that of graphene. By a 90-degree rotation, we get the Fermi-velocity of Weyl point P$_{2}$ (spin-down channel). The maximum of the velocity is in the y-direction, while the minimum is in the x-direction. Thus in the x-direction, the Fermi velocity in the spin-down channel is about three times the value in the spin-up channel, and in the y-direction, the Fermi velocity in the spin-up channel is about three times the value in the spin-down channel, illustrating a highly spin-momentum locked transport for spin-polarized electrons.

One important feature of Weyl points is that when a pair of Weyl points with opposite chirality are brought together, they will mutually annihilate \cite{Lu2013,PhysRevLett.121.106402}. Thus we can utilize this property to open up a gap at the Fermi level and transform AFM s-CrO into a semiconductor. As shown in Fig. \ref{figure4}, with equal-biaxial tensile strain (equal in the x- and y-directions) ranging from $\epsilon=0\%$ to $\epsilon=3\%$, the Weyl points P$_{1}$ and P$_{1}^{\prime}$ move towards X point and meet there. With larger equal-biaxial strain, the Weyl points annihilate, and a gap is opened up in the spin-down channel. Without symmetry broken, we can imply a gap is also opened up in the spin-up channel. From Fig. \ref{figure4} (a), we can observe that the Fermi velocities of the Weyl points in the spin-down channel are decreasing rapidly in the y-direction. In contrast, the Fermi velocities in the x-directions are not changed a lot. Therefore, with equal-biaxial tensile strain, we can increase the polarization of the transport properties in semimetallic AFM s-CrO.

With equal-biaxial strain larger than 3$\%$, AFM s-CrO becomes a semiconductor. In the y-direction in the spin-down channel, as shown in Fig. \ref{figure4} (c) up-panel, the VBM and CBM both become quadratic like a typical semiconductor. Interestingly, in the x-direction, as shown in Fig. \ref{figure4} (c) down-panel, the VBM and CBM are still close to linear dispersion, which gives rise to a ultra-high carrier mobility in the x-direction in the spin-down channel. The gap can be tuned with different equal-biaxial strain easily. In contrast, opening up a band-gap in graphene while preserving the linear dispersion is very attractive but challenging. In conclusion, in the semiconducting phase of AFM s-CrO, carriers with spin-down polarization prefer to move in the x-direction, while carriers with spin-up polarization prefer to move in the y-direction. 

With different unequal-biaxial strains (unequal in the x- and y-directions), we can get several states with spin-polarized transport properties. Among them, two types of spintronic states are the most interesting. One interesting spintronic state is the half-metallic AFM state with a pair of Weyl points in the spin-up or spin-down channel. For instance, as shown in Fig. \ref{figure4}(d), with an unequal-biaxial strain of $(0\%, 9\%)$, the Weyl points in spin-down channel P$_{1}$ and P$_{1}^{\prime}$ annihilate, while the Weyl points in spin-up channel $P_{2}$ and $P_{2}^{\prime}$ are preserved, giving rise to a half-metallic transport. Though the 90-degree rotation symmetry is broken, the net magnetism is still zero since the Weyl points are at the Fermi level. Another crucial kind of spintronic states is the half-semiconducting AFM states. With different strains in the x- and y-directions, for example, with a strain of $(3\%, 6\%)$, one can open up different band-gaps in two spin-channel, which means that the VBM and CBM are in the same spin-channel, giving rise to a half-semiconducting transport, as shown in Fig. \ref{figure4} (e). The net magnetism is still zero, with no bands passing through the Fermi level.

When unequal-biaxial strains are applied, the magnetic moments of the two Cr atoms are not equal, which means that they cannot cancel each other when we compute the total magnetic moment. The zero net magnetism in the above two spintronic states is originated from the spin-polarization of O atoms, which is discussed in Part VIII of the SM \cite{SuppMater}.

The novel spintronic properties of AFM s-CrO make it a promising material for two types of spin torques. As shown in Fig. S9 (a) \cite{SuppMater}, with charge current passing through s-CrO along the (110)-direction, the spin-up electrons prefer to move along the (010)-direction, while the spin-down electrons prefer to move along the (100)-direction. Thus spin-up and spin-down electrons accumulate at opposite edges, resulting in a spin current, which is usually achieved by SOTs. As shown in Fig. S9 (b), (c), and (d) \cite{SuppMater}, with charge current passing through s-CrO along the (100)-direction, the spin-down electrons are favored in the transport, which makes it indeed a STT. Moreover, the spin-channel of the STT can be switched by external strain.

Moreover, we have computed 240 different monolayer AB (A = Co, Cr, Fe, Mn, Mo, Nb, Ni, Pd, Rh, Ru, Sc, Ti, V, Y, Zr; B = B, C, N, O, F, Al, Si, P, S, Cl, Ga, Ge, As, Se, Br) with the same structure and AFM1 magnetic configuration. It is observed that among them, all the magnetic structures have giant spin-splitting in their band structures.

In summary, by utilizing first-principles calculations, we predicted a two-dimensional AFM spintronic material, s-CrO, with giant spin-splitting in the absence of spin-orbit coupling. AFM s-CrO turns out to be stable in both structural and magnetic configuration. It has a novel electronic band structure with two pairs of spin-polarized Weyl points in two spin channels at the Fermi level. With SOC, the Weyl points are preserved. The spin-down polarized Weyl points have their highest Fermi velocity in the x-direction, while the spin-up polarized Weyl points have their highest Fermi velocity in the y-direction. Under equal tensile strain in both x- and y-directions, the Weyl points can move towards the X (Y) points with their transport properties becoming more spin-polarized than in the pristine structure. After the Weyl points meet and mutually annihilate, a band-gap opens up while the linear dispersion in a specific direction is preserved, giving rise to semiconducting states with spin-polarized transport properties. More interestingly, with unequal biaxial strain, one can open up a gap in only one STC while preserving the Weyl points in the other spin-channel and get a half-metallic AFM semimetal. With larger unequal strain, one can open up different gaps in two STC and get AFM half-semiconducting states. Finally, we discussed the possible SOT-like and STT-like spin torques constructed by AFM s-CrO. We hope that this work will open up new avenues for collinear AFM spintronics.

We thank Mingwen Zhao, Raquel Esteban-Puyuelo and Suhas Nahas for helpful discussions. This work is supported by the project grant (2016-05366) and Swedish Research Links programme grant (2017-05447) from Swedish Research Council. Linyang Li acknowledges financial support from the National Natural Science Foundation of China (Grant No. 12004097), the Natural Science Foundation of Hebei Province (Grant No. A2020202031), and the Foundation for Introduction of Overseas Scholars of Hebei Province (Grant No. C20200313). Xin Chen and Duo Wang thank China scholarship council for financial support (No. 201606220031 and No. 201706210084). We also acknowledge SNIC-UPPMAX, SNIC-HPC2N and SNIC-NSC centers under the Swedish National Infrastructure for Computing (SNIC) resources for the allocation of time in high-performance supercomputers. Moreover, supercomputing resources from PRACE DECI-15 project DYNAMAT are gratefully acknowledged.

\bibliography{Ref}

\end{document}